\documentclass[twocolumn,superscriptaddress,amscd,amssymb,verbatim]{revtex4}

\usepackage{graphicx}
\usepackage{textcomp}
\usepackage[OT1,T1]{fontenc}
\usepackage{color}
\definecolor{red}{rgb}{1,0,0}

\usepackage{epstopdf}
\usepackage{bm}  

\begin{document}
\newcommand{\alfa}{$\alpha$-NaMnO$_2$}

\title{Frustration-induced nanometre-scale inhomogeneity in a triangular antiferromagnet}
\author{A. Zorko}
\email{andrej.zorko@ijs.si}
\affiliation{Jo\v{z}ef Stefan Institute, Jamova c.~39, 1000 Ljubljana, Slovenia}
\affiliation{EN--FIST Centre of Excellence, Dunajska c.~156, SI-1000 Ljubljana,
Slovenia}
\author{O. Adamopoulos}
\affiliation{Institute of Electronic Structure and Laser, Foundation
for Research and Technology -- Hellas, Vassilika Vouton, 71110 Heraklion, Greece}
\author{M. Komelj}
\affiliation{Jo\v{z}ef Stefan Institute, Jamova c.~39, 1000 Ljubljana, Slovenia}
\author{D. Ar\v{c}on}
\affiliation{Jo\v{z}ef Stefan Institute, Jamova c.~39, 1000 Ljubljana, Slovenia}
\affiliation{Faculty of Mathematics and Physics, University of Ljubljana, Jadranska c.~19,
1000 Ljubljana, Slovenia}
\author{A. Lappas}
\affiliation{Institute of Electronic Structure and Laser, Foundation
for Research and Technology -- Hellas, Vassilika Vouton, 71110 Heraklion, Greece}

\date{\today}
\begin{abstract}
Phase inhomogeneity of otherwise chemically homogenous electronic systems is an essential ingredient leading to fascinating functional properties, such as high-$T_c$ superconductivity in cuprates, colossal magnetoresistance in manganites, and giant electrostriction in relaxors. In these materials distinct phases compete and can coexist due to intertwined ordered parameters. Charge degrees of freedom play a fundamental role, although phase-separated ground states have been envisioned theoretically also for pure spin systems with geometrical frustration that serves as a source of phase competition. Here we report a paradigmatic magnetostructurally inhomogenous ground state of the geometrically frustrated $\alpha$-NaMnO$_2$ that stems from the system's aspiration to remove magnetic degeneracy and is possible only due to the existence of near-degenerate crystal structures. Synchrotron X-ray diffraction, nuclear magnetic resonance and muon spin relaxation show that the spin configuration of a monoclinic phase is disrupted by magnetically short-range ordered nanoscale triclinic regions, thus revealing a novel complex state of matter. 
\end{abstract}

\maketitle

Complexity of phases is a common theme in systems characterized by simultaneously active degrees of freedom -- spin, charge, orbital or phonon \cite{Dagotto}. 
Even when Hamiltonians possess translational symmetry this may be broken on a nano- or a mesoscale owing to competing electronic states \cite{Dagotto,Shenoy}. Such inhomogeneities appear in various oxides, including cuprates \cite{Tranquada,Hanaguri,Vojta}, manganites \cite{DagottoPR,Loudon, Burkhardt}, cobaltates \cite{Roger}, nickelates \cite{TranquadaPRL,Wawrzynska}, ferrites \cite{Wu,Groot} and iron pnictides \cite{Park,LangPRL}. In these systems inhomogeneity is related to the charge degrees of freedom and usually appears because of quenched disorder -- random deviations from perfect system uniformity -- which can dramatically alter the material's properties \cite{Dagotto}. 
Such a disorder is also held responsible for a nanoscale magnetoelastic texturing that appears in the absence of charge degrees of freedom in a magnetically ordered state of certain localized spin systems \cite{Sexana}. This inhomogeneity is only a precursor of a displacive structural transition and is limited to a certain temperature ($T$) interval. On the other hand, inhomogenous magnetic ground states at $T\rightarrow 0$ are rare and poorly understood \cite{ZorkoCuNCN,Nakajima}, although they have been predicted to appear already in a clean limit (no quenched disorder)
of frustrated magnets \cite{Schmalian, Mu, Kamiya}.
Geometrical frustration of a spin lattice \cite{Misguich,LMM} is known for provoking novel fundamental phenomena in magnetism, like exotic disordered ground states and fractional spin excitations \cite{LMM,Balents,Han}. Moreover, in improper multiferroics it leads to uniform lattice distortions  \cite{Cheong}.
Therefore, ground-state degeneracy of frustrated magnets, in alliance with magnetoelastic coupling, could make the energetic cost of spatial fluctuations low, rendering them "electronically soft" \cite{Milward}. These systems are thus perfect candidates for yet unseen ground-state inhomogeneities, emerging solely in the magnetostructural channel. 

\begin{figure*}[t]
\includegraphics[trim = 1mm 1mm 1mm 1mm, clip, width=0.689\linewidth]{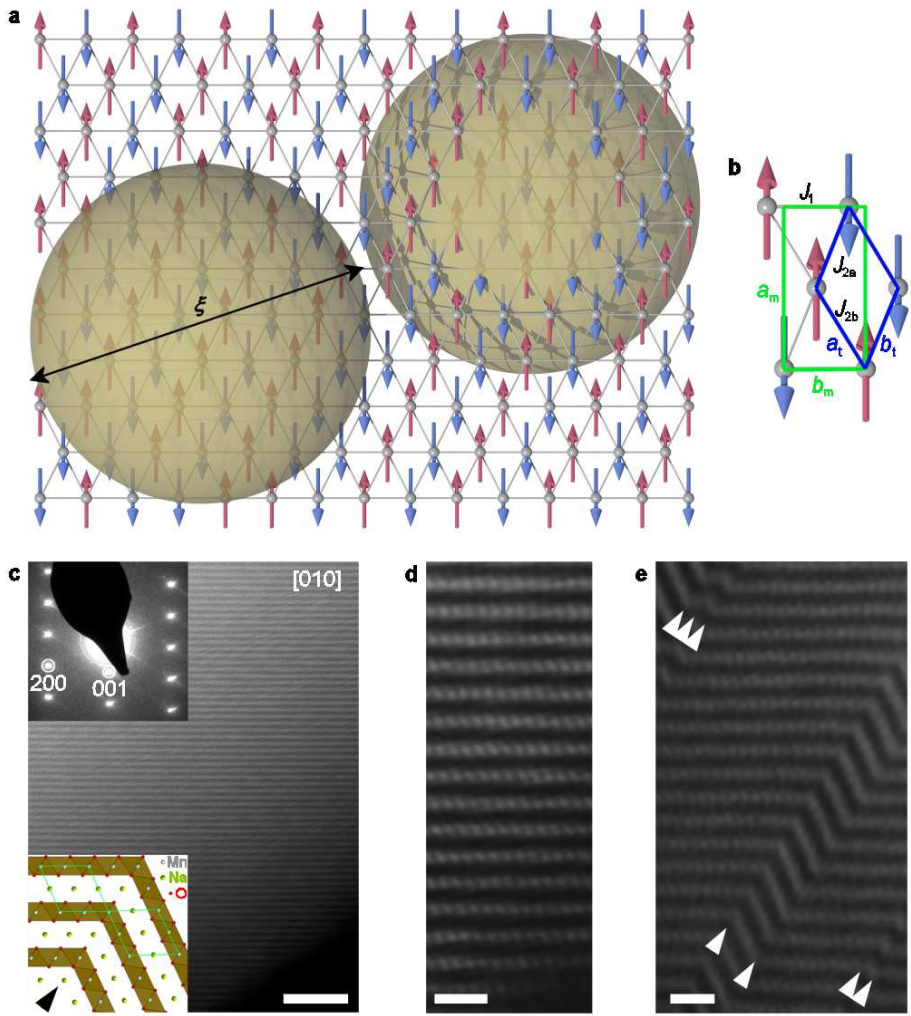}
\caption{{\bf Structure of \alfa.} {\bf (a)} Anisotropic triangular lattice of $S=2$ (Mn$^{3+}$) spins (arrows) in \alfa~in a selected $ab$ crystallographic plane of the monoclinic (m) cell settings. Long-range antiferromagnetic order with wave vector ${\bf k} =(\frac{1}{2},\frac{1}{2},0)$ is interrupted inside nano-sized magnetostructural defect region (represented by spheres with radius $\xi$) of triclinic crystal symmetry (t). {\bf (b)} Schematic view of the lattice distortion in the triclinic phase with respect to the monoclinic phase. The Mn-Mn  chains with the dominant intrachain antiferromagnetic exchange $J_1$ shift along the $b_{\rm m}$ axis, making the two interchain constants $J_2$ distinct ($J_{2{\rm a}} > J_{2{\rm b}}$ ). {\bf (c-e)} Room-temperature HAADF-STEM images of [010] planes in \alfa~with rows of bright dots representing Mn$^{3+}$ ions stacked along the $a_{\rm m}$ crystallographic axis. Both, {\bf (c)} large regions of faultless monoclinic structure and {\bf (e)} areas of crystal faults with twin boundaries (TBs), marked by arrows, confined to the (-101) lattice plane are found. TBs tend to coalesce forming more complex TBs, consisting of odd number of individual TBs, or stacking faults (SFs), consisting of even number of individual TBs. A model TB and a diffraction image are shown in the insets to {\bf c}. The horizontal scale bar in {\bf (c)} corresponds to 5~nm and in {\bf (d,e)} to 1~nm.}
\label{fig1}
\end{figure*}

The layered $\alpha$-NaMnO$_2$ compound may possess the required properties of such a state. It represents a spatially anisotropic triangular lattice of $S = 2$ (Mn$^{3+}$) spins with dominant intrachain ($J_1 =65$~K) and frustrated interchain ($J_2 =29~$K) antiferromagnetic exchange interactions \cite{Zorko} in the $ab$ plane (Fig.~\ref{fig1}). The ground state is dictated by strong single-ion anisotropy \cite{Zorko} and the frustration renders the magnetic excitations one dimensional (1D) \cite{Stock}. Both sharp magnetic reflections and diffuse scattering appearing at the same wave vector were simultaneously observed in neutron powder diffraction (NPD), even far below the N\'eel temperature $T_{\rm N}=45$~K \cite{Giot}. Such a coexistence of long- and short-range spin correlations indicates a magnetically inhomogenous ground state. Moreover, large and highly anisotropic microstrains in the crystal structure above $T_{\rm N}$ suggest structural inhomogeneities. They can be thought of as a precursor of the structural phase transition from a monoclinic to a triclinic symmetry, which was previously argued to occur concurrently with the magnetic transition due to magnetoelastic coupling \cite{Giot}. However, no direct evidence for the structural change, marked by splitting of nuclear Bragg peaks, has so far been identified. Alternatively, all the observed magnetic and structural peculiarities of \alfa~could be fingerprints of an intrinsically inhomogenous ground state.
Such an "electronically soft" state could be related to favourable electrochemical properties of \alfa, characterized by its relatively high discharge capacity and structural stability found after cycling \cite{Qu,Ma,Kim}.

In this work, we have utilized high-resolution synchrotron X-ray diffraction (XRD) combined with the atomic-resolution high-angle annular dark-field scanning transmission electron microscopy (HAADF-STEM) to identify previously unseen subtle structural modifications of \alfa. Furthermore, we demonstrate by means of complementary local-probe nuclear magnetic resonance (NMR) and muon-spin relaxation ($\mu$SR) methods that a single-phase stoichiometric \alfa~sample adopts a unique inhomogenous ground state. We show that this is an outcome of the geometrical frustration and simultaneously active spin and lattice degrees of freedom, which through a lattice-distorting procedure, endorsed by quasi-elastic magnetic excitations, lead to a spontaneous emergence of competing magnetostructural states at the nanoscale. This novel paradigm in the context of frustration-driven phase complexity is of conceptual importance in condensed matter physics. It allows better understanding of the fundamental mechanisms that are responsible for a plethora of intriguing quantum phenomena arising from degenerate ground states.

\section*{Results}
\begin{figure*}[t]
\includegraphics[trim = 6mm 62mm 4mm 20mm, clip, width=0.8\linewidth]{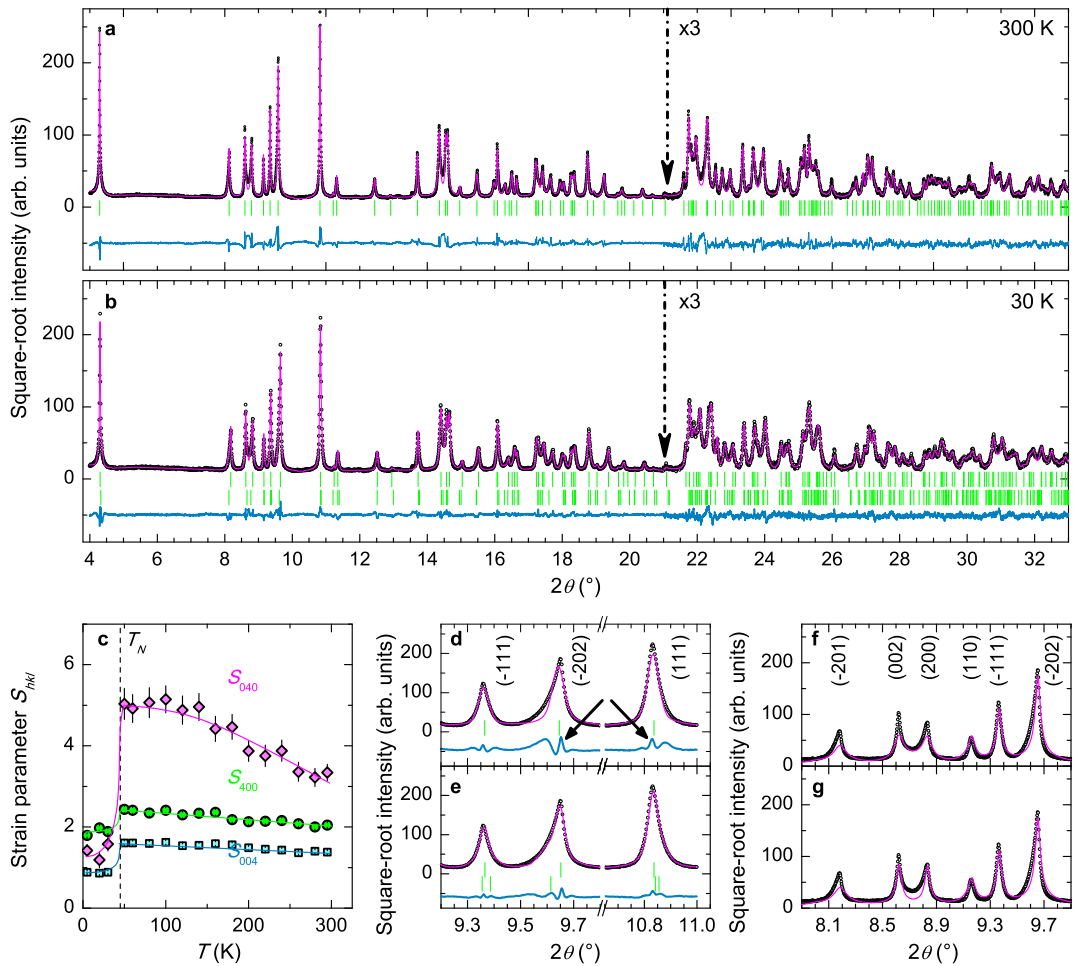}
\caption{{\bf Synchrotron X-ray scattering.} {\bf (a)} Observed (circles), calculated (thin line) and difference (thick line) synchrotron XRD profile of \alfa~at 300~K, with vertical ticks as predicted reflections, after Rietveld refinement assuming the single monoclinic phase [$C2/m$; $a= 5.67075(5)$~\AA, $b= 2.85896(2)$~\AA, $c= 5.80109(6)$~\AA, $\beta= 113.1537(4)^\circ$]. {\bf (b)} The 30~K profile, after Rietveld refinement with
the two-phase model of coexisting monoclinic [$C2/m$; $a= 5.63750(3)$~\AA, $b= 2.85763(1)$~\AA, $c= 5.77295(3)$~\AA, $\beta= 112.8926(4)^\circ$] and triclinic [$P\bar{1}$; $a= 3.1814(2)$~\AA, $b= 3.1751(2)$~\AA, $c= 5.7589(2)$~\AA, $\alpha= 110.429(3)^\circ$, $\beta= 110.589(4)^\circ$, $\gamma= 53.308(3)^\circ$] phases (see Methods). At $2\theta> 21^\circ$ intensities are multiplied by 3 to improve clarity.
{\bf (c)} The temperature evolution of the anisotropic strain-broadening parameters $S_{hkl}$ (lines are a guide to the eye). {\bf (d-g)} The details of the 30~K profile after Rietveld fits with {\bf (d)} the monoclinic phase (arrows show important deviations) and {\bf (e)} coexisting strained monoclinic and unstrained triclinic phases.  Comparison of the same profile with a calculated pattern assuming 2\% (in volume) stochastically distributed {\bf (f)} twin boundaries and {\bf (g)} stacking faults (see Methods). arb. units, arbitrary units. The error bars in the diffraction patterns are defined as a square root of the counted photons, while those of the $S_{hkl}$ parameters are standard deviations of the fit parameters in the Rietveld refinements.}
\label{fig2}
\end{figure*}
\noindent {\bf Structural complexity.}
A new high-resolution insight into the large strain fields detected in the early NPD study \cite{Giot} is provided by the synchrotron XRD measurements shown in Fig.~\ref{fig2}. Above $T_{\rm N}$ the monoclinic structure $C2/m$, a variant of the prototype rhombohedral delafossite due to the presence of Jahn-Teller active Mn$^{3+}$ cations, is confirmed (see Fig.~\ref{fig2}a) if $hkl$-dependent anisotropic broadening of the Bragg reflections is taken into account. Within the Stephens's formalism \cite{Stephens} of anisotropic strain broadening (strain broadening parameters $S_{hkl}$) we find that the $S_{040}$ parameter is the largest and increases the most down to $T_{\rm N}$ (Fig.~\ref{fig2}c). This suggests that the distribution of the $ac$-interplane distances is the broadest and highly affected when approaching $T_{\rm N}$ from above, while the $ab$ ($S_{004}$) and $bc$ ($S_{400}$) interplane-distance distributions are much narrower and less temperature dependent. Qualitatively, the $S_{040}$ strains in the monoclinic structure may originate from minute shifts of the Mn-Mn spin-chains along the $b_{\rm m}$ axis (Fig.~\ref{fig1}b), a lattice distortion that relieves the frustration and could effectively lead to the previously proposed $P \bar 1$ triclinic structure at $T< T_{\rm N}$ \cite{Giot}.
\begin{figure*}[t]
\includegraphics[trim = 2mm 50mm 3mm 15mm, clip, width=0.8\linewidth]{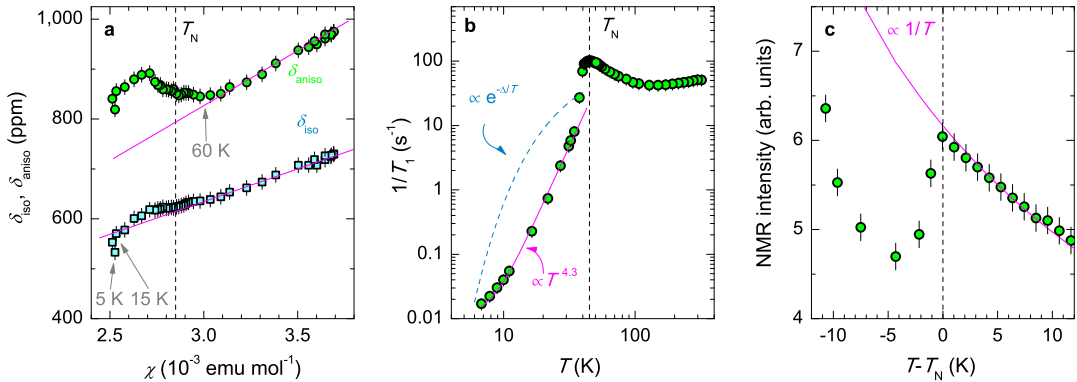}
\caption{{\bf $^{23}$Na NMR evidence of inhomogenous magnetic state below $T_{\rm N}$.} {\bf (a)} Scaling of the central transition ($-\frac{1}{2}\leftrightarrow \frac{1}{2}$) $^{23}$Na NMR line shift (isotropic $\delta_{\rm iso}$ hyperfine coupling) and line width (anisotropic $\delta_{\rm aniso}$ hyperfine coupling) with bulk susceptibility $\chi $ at 5~T, and thus implicitly with temperature. Solid lines are linear fits of the high-$T$ data. {\bf (b)} The $^{23}$Na spin-lattice relaxation rate $1/T_1$ exhibiting a power-law dependence below $T_{\rm N}$ (the additional low-$T$ constant term $1/T_1^0=0.01$~s$^{-1}$ corresponding to residual relaxation). {\bf (c)} Partial wipeout of the central NMR signal (corrected for spin-spin relaxation) below $T_{\rm N}$. All NMR data were acquired at the fixed applied field of 8.9~T. ppm, parts per million; arb. units, arbitrary units. Error bars represent standard deviations of the fit parameters (see Methods).}
\label{fig3}
\end{figure*}

In contrast to the isostructural CuMnO$_2$ \cite{Vecchini, Poienar}, the high-resolution synchrotron XRD patterns of \nolinebreak{$\alpha$-NaMnO$_2$} below $T_{\rm N}$, similar to NPD, do not show any resolvable peak splitting that would unambiguously confirm a bulk phase transformation to the lower-symmetry triclinic phase (see Methods). Surprisingly, Rietveld refinements at $T< T_{\rm N}$ on the highly parametrised, strained triclinic phase (t) produce a statistically poorer quality of fit compared to the strained monoclinic (m) phase, $\chi^2_{\rm t}\approx 1.4\chi^2_{\rm m}$ (see Methods). A close inspection of the fitted patterns reveals asymmetrically broadened profiles in terms of the peak shape, where the calculated pattern systematically misses intensity from the low- and high-angle tails of certain Bragg reflections (Fig.~\ref{fig2}d). Given the intrinsic resolution difference between NPD and XRD, such a pattern complexity could not have been revealed by the early NPD (see Methods) that was fitted satisfactorily by the triclinic model \cite{Giot}. The synchrotron XRD data, therefore, suggests that a more complex parametrization of the structural model is required.

Presence of stacking faults \cite{Croguennec} or structural inhomogeneities \cite{Lappas} in otherwise single-phase materials, may be the reason for the development of short-scale regions, which can give rise to unusually broadened profiles, like those of \nolinebreak{$\alpha$-NaMnO$_2$}. To inspect these, HAADF-STEM was used at room temperature. Large faultless regions that give rise to distinct diffraction spots indexed in the $C2/m$ symmetry, are found (Fig.~\ref{fig1}c,d). However, these are broken by planar defects that entail coherent mirror twins separated by twin boundaries and stacking faults (Fig.~\ref{fig1}e). Such planar defects are not expected to evolve notably with decreasing temperature, while, in contrast, the $S_{040}$ strain broadening parameter increases significantly from room temperature down to $T_{\rm N}$ (Fig.~\ref{fig2}c). Simulations of the diffraction profile resulting from such defects (see Methods) yield anisotropic peak broadening (Fig.~\ref{fig2}f,g) that could partially contribute to the 30~K experimentally observed $hkl$-dependent anisotropic broadening of Bragg reflections. However, the simulations show that defects of both types fail completely to explain the strongly asymmetric peak shape of certain Bragg reflections, for example, $(-202)$, and are thus incapable of describing the synchrotron XRD data below $T_{\rm N}$.
 
A way of simulating such a behaviour is by a two-phase (2p) Rietveld model accounting for a nanoscale coexistence of the monoclinic and the triclinic phases. This model provides significantly improved quality of fit, $\chi^2_{\rm 2p}\approx 0.35\chi^2_{\rm m}$ (Fig.~\ref{fig2}b,e). The obtained volume fraction of the triclinic phase is $\sim$35(1)$\%$ at 30~K and only marginally evolves down to 5~K. The majority monoclinic phase retains some strain, but this is severely relieved below $T_{\rm N}$ (Fig.~\ref{fig2}c), where the  two phases coexists. Randomly distributed nanoscale inhomogeneities, pertaining to the minority triclinic phase that is grown at the expense of the majority monoclinic phase, may severely modify the electronic response (\textit{vide infra}) of the bulk compound. This renders it an intriguing subject for magnetic local-probe investigations.  
\begin{figure*}[t]
\includegraphics[trim = 6mm 15mm 4mm 5mm, clip, width=0.78\linewidth]{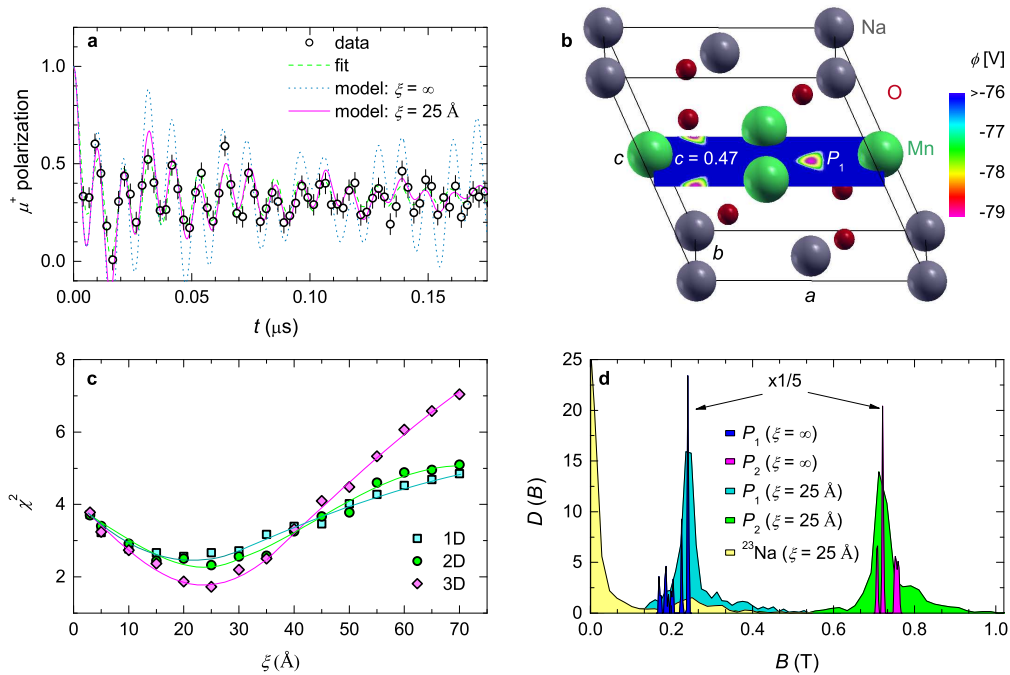}
\caption{{\bf $\mu$SR determination of local-magnetic-field distributions in the inhomogenous magnetic phase.} {\bf (a)} Experimental muon polarization at 5~K (circles), compared to the modelling with nanoscopic inhomogeneities (see Methods) and fitting with two damped cosine contributions, 
$P(t) = \sum_{j=1}^{2}  {f_j\left[\frac{1}{3}{\rm e}^{-\lambda_{\rm l} t}+\frac{2}{3}{\rm cos}(\gamma_\mu B_{\mu,j}){\rm e}^{-\lambda_{{\rm t},j} t}\right]}$, where $f_j$ denotes muon occupation probabilities for the two stopping sizes $j$, $B_{\mu,j}$ the local field, $\lambda_{\rm l}$ the longitudinal and $\lambda_{{\rm t},j}$ the transversal muon relaxation rates. {\bf (b)} Contour of the electrostatic potential $\phi$ in the $ab$ plane within the \alfa~unit cell obtained by {\it ab initio} calculations (see Methods) with highlighted global minima at the $P_1=(0.18,0,0.47)$ site. {\bf (c)} The reduced $\chi^2$ of modelling the 5~K $\mu$SR polarization curve with the models of one-, two- and three-dimensional magnetic defects with concentration $p=0.35$, all yielding the same optimal correlation length $\xi=25$\AA~(lines are a guide to the eye). {\bf (d)} The calculated field distributions on both muon stopping sites, $P_1$ and $P_2$, and on the $^{23}$Na nuclear site for the one-dimensional-defect model with $\xi=\infty$ and the model of three-dimensional finite-size defects ($\xi=25$~\AA) (see Methods).  The error bars of muon-polarization data are defined as a square root of the total number of detected positrons resulting from muon decays.}
\label{fig4}
\end{figure*}

\vspace{0.5 cm}
\noindent  {\bf Magnetic inhomogeneity revealed by NMR.}
To validate the apparent phase separation and associated local inhomogeneities in \alfa, we employed the NMR technique, locally probing electronic charge and magnetic environment of the $^{23}$Na ($I=3/2$) nuclei through their quadrupolar coupling with the electric field gradient and hyperfine coupling with the electronic spins, respectively. The scaling of the central transition ($-\frac{1}{2}\leftrightarrow \frac{1}{2}$) NMR line shift $\delta_{\rm iso}$ (due to isotopic hyperfine coupling) and the line width $\delta_{\rm aniso}$ (due to the anisotropic hyperfine coupling) with bulk susceptibility $\chi$ is shown in Fig.~\ref{fig3}a. They both originate from local static hyperfine magnetic fields. Therefore, they are expected to scale linearly with $\chi$, not only in the paramagnetic state but also below $T_{\rm N}$, because the local fields from  antiferromagnetically ordered Mn$^{3+}$ moments at sites $\pm {\bf r}$ \cite{Giot} effectively cancel out at the $^{23}$Na position with $2/m$ site symmetry. The apparent low-temperature non-linearity $\delta_{\rm aniso}\not\propto \chi$ that is not due to quadrupolar-broadening effects (see Methods) discloses a more complicated magnetic order than initially proposed \cite{Giot}. We stress that $\delta_{\rm aniso}$ does not exhibit an order-parameter-like dependence below $T_{\rm N}$; hence, it cannot be explained by a small additional spin component that would break the symmetry at the $^{23}$Na sites. 

A sudden suppression of the $^{23}$Na NMR spin-lattice relaxation rate $1/T_1$ at $T_{\rm N}$ shown in Fig.~\ref{fig3}b proves that NMR is sensitive to low-energy magnetic excitations and that the magnetic transition is sharp and occurs instantaneously throughout the sample.
We find that the stretch exponent of the nuclear-magnetization recovery curves in the $T_1$ experiment (see Methods) that is constant, $\beta=0.86(2)$, at $T>T_{\rm N}$ dramatically decreases in the magnetically ordered state and reaches the value $\beta=0.60(2)$ at lowest temperatures. This is a fingerprint of an increased distribution width of $1/T_1$, and hence a broadened distribution of local environments, signalling an inhomogenous magnetic ground state of \alfa.

The inhomogenous magnetic ground state is further supported by a sudden drop of the NMR intensity below $T_{\rm N}$ by $\sim 30\%$ (Fig.~\ref{fig3}c). As the NMR intensity should normally scale with $1/T$ due to the Boltzmann population factor, this partial NMR wipeout can only be attributed to an unobservable NMR component accompanying the remaining NMR component that is characterized by a zero static local magnetic field. NMR thus clearly discloses a magnetic phase separation below $T_{\rm N}$.

\vspace{0.5 cm}
\noindent {\bf $\mu$SR detection of magnetic inhomogeneity.}
To expose the microscopic nature of the magnetic phase separation below $T_{\rm N}$, we finally resort to $\mu$SR that is extremely sensitive to local magnetic fields $B_\mu$, and directly measures their distribution in an ordered state through the relaxation function $P(t)$ of the muon polarization \cite{Yaouanc}. Below $T_{\rm N}$, levelling of $P(t)$ at $1/3$ for longer times (Fig.~\ref{fig4}a) that is characteristic for powder samples with randomly oriented internal magnetic field with respect to initial muon polarization, is a clear indication that the bulk of the sample has undergone a transition into the ordered state. Early-time oscillations, on the other hand, reveal that two inequivalent muon stopping sites $P_1$ and $P_2$ exist. The occupation probabilities for the two sites $f_1$:$f_2=60(5)$:$40(5)$ are temperature independent and the corresponding internal fields at 2~K are $B_{\mu 1}=0.22(2)$~mT and $B_{\mu 2}=0.69(2)$~mT. 

We emphasize that the satisfactory fitting of the experimental muon polarization curve recorded at 5~K to a model of two damped cosine contributions (Fig.~\ref{fig4}a) reveals a large difference between the longitudinal $\lambda_{\rm l} =0.03(1)$~\textmu s$^{-1}$ and the transverse $\lambda_{\rm t,1} =82(5)$~\textmu s$^{-1}$, $\lambda_{\rm t,2} =12(1)$~\textmu s$^{-1}$ muon relaxation rates. The former is of dynamical origin, while the latter additionally encompasses a distribution of static local fields \cite{Yaouanc}. This distribution is obviously responsible for the fast damping of oscillations and must originate from the inhomogenous magnetic state. To model it, the knowledge of muon stopping sites is required. Here, we rely on \textit{ab initio} calculations (see Methods), which predict the two most likely muon stopping sites $P_1=(0.18,0,0.47)$ and $P_2=(0.45,0,0.75)$; the former corresponds to the global electrostatic-potential minimum (Fig.~\ref{fig4}b). The calculated dipolar fields (see Methods) from antiferromagnetically ordered Mn$^{3+}$ moments at these sites match the experimentally determined fields $B_{\mu 1}$ and $B_{\mu 2}$.

\section*{Discussion}
The broad local-magnetic-field distributions at the muon stopping sites $P_1$ and $P_2$ provide a microscopic insight to the inhomogenous magnetic state of \alfa. Most importantly, they can be related to the "parasitic" triclinic structural phase suggested by the synchrotron XRD. We could model (see Methods) these field distributions by introducing triclinic magnetic defects in the form of dispersed nano-sized islands with the magnetic order reversed at the interfaces with the monoclinic matrix, as schematically presented in Fig.~\ref{fig1}a. Although the domain walls in magnets tend to be chiral and possess a finite width \cite{Catalan}, we envisage that this width will be small in \alfa, because of its rather large magnetic anisotropy \cite{Zorko}. Therefore, the sketch of the inhomogenous ground state in Fig.~\ref{fig1}a should be a good approximation of the actual situation.

We have considered the magnetic defect regions to be either one- (1D), two- (2D), or three-dimensional (3D) objects with a characteristic size $\xi$, randomly positioned in the lattice so that a given fraction $p$ of Mn$^{3+}$ ions resides in them. Independent on their dimensionality, when $\xi = \infty $ (corresponding to stripe- and slab-shifts for 1D and 2D magnetic defects, respectively, as well as for the observed structural twin boundaries and stacking faults) we find narrow local-field distributions with only a few well-defined frequencies (Fig.~\ref{fig4}d). These are manifested in beating (Fig.~\ref{fig4}a) rather than in experimentally observed damping of $P(t)$. On the other hand, for finite $\xi$ the field distributions become much broader (Fig.~\ref{fig4}d) and, consequently, result in heavily damped $P(t)$, in accord with the experiment. For $p=0.35$, as derived from the structural analysis, we can reproduce the experimental $P(t)$ at 5~K rather well with $\xi=25(5)$\AA~(Fig.~\ref{fig4}a), independent of the defect dimensionality, though the 3D model yields the best agreement with the experiment (Fig.~\ref{fig4}c). The agreement of our modelling with the experiment is very good, bearing in mind that $\xi$ is the only free parameter once the parameter $p$ is fixed to the value inferred from XRD. Even better quality of modelling could possibly be achieved by further complicating the model with a distribution of defect sizes $\xi$ or a finite width of phase boundaries (domain walls). However, at this point, such complications seem unnecessary, as our approach already captures the main features of $\mu$SR polarization decay and thus unambiguously demonstrates the presence of nanoscopic magnetic domains in \alfa, with well-defined average size.  

The dispersed finite-size defects are in conformity with the short-range ordered regions inferred by neutron diffraction \cite{Giot}. We note that the correlation length deduced from the neutron diffraction experiments at 5~K appears somewhat smaller; however, it is likely underestimated considering the amount of diffuse scattering at this temperature. On the other hand, $\xi=25(5)$\AA~is in good accordance with the correlation length derived from the diffuse neutron peak in the temperature range between 15~K and $T_{\rm N}$ \cite{Giot}.
Moreover, introduction of magnetostructural defects also explains the sudden $\sim$30\% decrease of the NMR intensity below $T_{\rm N}$ (Fig.~\ref{fig3}c). Namely, the defects locally break the inversion symmetry and lead to a two-peak local-field distribution at the $^{23}$Na sites (Fig.~\ref{fig4}d). The dominant peak remains positioned at $B_{{\rm Na}1} = 0$ and is modestly broadened, which may explain the low-temperature deviation of $\delta_{\rm aniso}$ from linearity (Fig.~\ref{fig3}a). The second peak is found at $B_{{\rm Na}2} = 0.26$~T, yielding a huge width of the corresponding NMR powder signal ($B_{{\rm Na}2}/B_0=3\%$; $B_0=8.9$~T is the applied magnetic field) that renders this component unobservable.

The parameter $\xi$ possesses no notable temperature dependence between 2~K and $\sim$30~K, while it slightly decreases at higher temperatures. Its temperature independence below $\sim$30~K clearly indicates that both magnetostructural phases are stable and, therefore, compete over a broad temperature range. We stress that the observed nanometer-scale inhomogeneity can neither occur solely at grain boundaries since the crystalline grains are several orders of magnitude larger, nor it can originate from the diluted planar structural defects whose topology is different. We rather argue that this unique state arises from the geometrical frustration of the underlying spin lattice, as a result of the competition between the elastic, magnetic-exchange and magnetoelastic energies. 
To support this hypothesis, we compare the total
energies of the relevant phases, calculated {\it ab initio} within
the framework of the density-functional theory (for technical details, see 
Methods). We considered the monoclinic and triclinic structures in either 
the nonmagnetic or antiferromagnetic states.  It is assumed that the 
theoretical nonmagnetic state is equivalent to the experimental paramagnetic state. The calculations were performed for theoretical equilibrium lattice parameters and relaxed atomic 
positions. Thus, any structural distortion for nonmagnetic states results in
the gain of the elastic energy, whereas the magnetic ordering yields 
the magnetic-exchange contribution as well as the magnetoelastic contribution.   
According to the results of these 
calculations, in the absence of magnetism the monoclinic phase is energetically 
more favourable than the triclinic one, in agreement with the experiment 
(Fig.~\ref{fig2}a). In the presence of the long-range antiferromagnetic
order the two magnetic contributions make the triclinic phase more favourable.
The estimated magneto-elastic contribution is of the order of
only a few ${\rm\mu eV}$ per unit cell (see Methods); hence the increase of the elastic 
energy due to the monoclinic-to-triclinic transformation, which is a few orders of magnitude larger, must be mainly balanced
by the magnetic exchange. The lowering of the magnetic-exchange energy is associated with the magnetic degeneracy being removed by the triclinic distortion, rendering the two interchain exchange constants alike (Fig.~\ref{fig1}b) by 0.2~meV \cite{Jia}.
The two opposing effects of similar magnitude lead to near degeneracy of the monoclinic and the triclinic phases \cite{Jia,Ouyang}, which enables already infinitesimal quenched disorder to locally lift the inherent frustration of the monoclinic phase. It thus provides a solid foundation for realization of a multiple-minima free-energy landscape that is required for formation of nanoscale domains. A similar situation is found in manganites \cite{Ahn}, but in contrast to \alfa, polaronic elastomagnetic textures in manganites are born from the coupling of elastic, spin and charge degrees of freedom \cite{Bishop} and are believed to be due to a martensitic accommodation strain \cite{Ahn}.

To better understand the mechanism stabilizing the remarkable inhomogenous ground state of \alfa, we inspect its low-energy magnetic excitations. According to inelastic neutron scattering, these should be gapped ($\Delta_0=87$~K) \cite{Stock}. Since these measurements can easily miss a quasi-elastic part, we resort to the complementary $^{23}$Na spin-lattice relaxation, which is governed by inelastic scattering of spin waves below $T_{\rm N}$ \cite{Moriya}. In the case of gapped excitations the thermally-activated relaxation rate $1/T_1 \propto {\rm e}^{-\Delta/T}$ is expected for $T\ll \Delta$, and the power-law dependence $1/T_1 \propto T^n$ for $T\gg \Delta$ \cite{Mila}. Here $n\geq 2$ depends on the dimensionality of the spin-wave dispersion and the number of magnons assisting the relaxation process \cite{Moriya,Mila}.  The expected exponential law with $\Delta=\Delta_0-\mu B_0=53$~K ($\mu=2.9\mu_{\rm B}$ \cite{Giot}) fails completely to fit the data (Fig.~\ref{fig3}b). In clear contrast, the power law with $n=4.3$  matches the experimental data very well, implying a very small $\Delta$, if any. Deviations from the power law are observed only in a critical region close to $T_{\rm N}$. Thus, our $1/T_1$ measurements clearly witness quasi-elastic magnetic excitations. Sliding of magnetic stripes along the $b_{\rm m}$ axis costs zero energy in the monoclinic setting due to the frustrated nature of the interchain exchange and is thus the main candidate for the $\Delta\rightarrow 0$ excitations. The individual stripes can shift independently and, therefore, the ensemble of such excitations becomes extended. As such excitations oppose a uniform magnetic state, they provide a likely rationalization of the observed unique magnetostructural inhomogeneity in \alfa.

Our present findings reach beyond superconducting cuprates, colossal magnetoresistant manganites and other chemically homogenous systems, which provide prominent examples of electronic-charge inhomogeneities that are mainly realized by means of tuning their chemical composition \cite{Dagotto,Shenoy,Tranquada,Hanaguri,Vojta,DagottoPR,Loudon,Burkhardt,Roger,TranquadaPRL,Wawrzynska,Park,LangPRL}. 
In stoichiometric \alfa~with no charge degrees of freedom, we have discovered an inhomogenous ground state solely in the magnetostructural channel. 
Therefore, we propose that the geometrical frustration on a spin lattice can lead to magnetostructural inhomogeneities, a new paradigm among phase-separated states of highly correlated electron systems. Such a state is likely to materialize in other frustrated magnets comprising near-degenerate crystal structures. 
Together with the recently discovered inhomogenous charge-cluster glass on a triangular charge lattice of a dopant-free organic conductor \cite{Kagawa}, this validates experimentally the geometrical frustration as a conceptionally novel mechanism that supports inhomogenous ground states, even in the absence of quenched disorder.

\section*{Methods}
\noindent {\bf Sample.}
All measurements were performed on the same high-quality polycrystalline sample that was used before \cite{Giot, Zorko, Stock}. The Rietveld analysis of the high intensity synchrotron XRD data found no deviation from the stoichiometric molecular type. If any impurity phases were present these were at a level below our detection capability of 0.3\% in volume. HAADF-STEM images confirmed the single-phase nature of the sample and witnessed large regions with good crystallinity. 

\vspace{0.5 cm}
\noindent {\bf Synchrotron XRD measurement.}
A thorough structural investigation of \alfa~in the temperature range of $T= 5-300$~K was performed by employing high-resolution synchrotron powder XRD in the ID31 beam-line ($\lambda= 0.39986$~\AA) at the European Synchrotron Radiation Facility, Grenoble, France. Our sample was sealed in an 1-mm borosilicate glass capillary, prepared inside a He(g)-filled glove-bag. This procedure permits for a tiny amount of He(g) trapped in the capillary to operate as an exchange gas that allows for effective thermalisation of the sample down to the base temperature (5~K), attainable with the continuous flow He-cryostat. Use of this thermal protocol provided excellent reproducibility of the XRD diffractograms on different batches. 
\begin{figure*}[t]
\includegraphics[trim = 0mm 22mm 0mm 25mm, clip, width=0.95\linewidth]{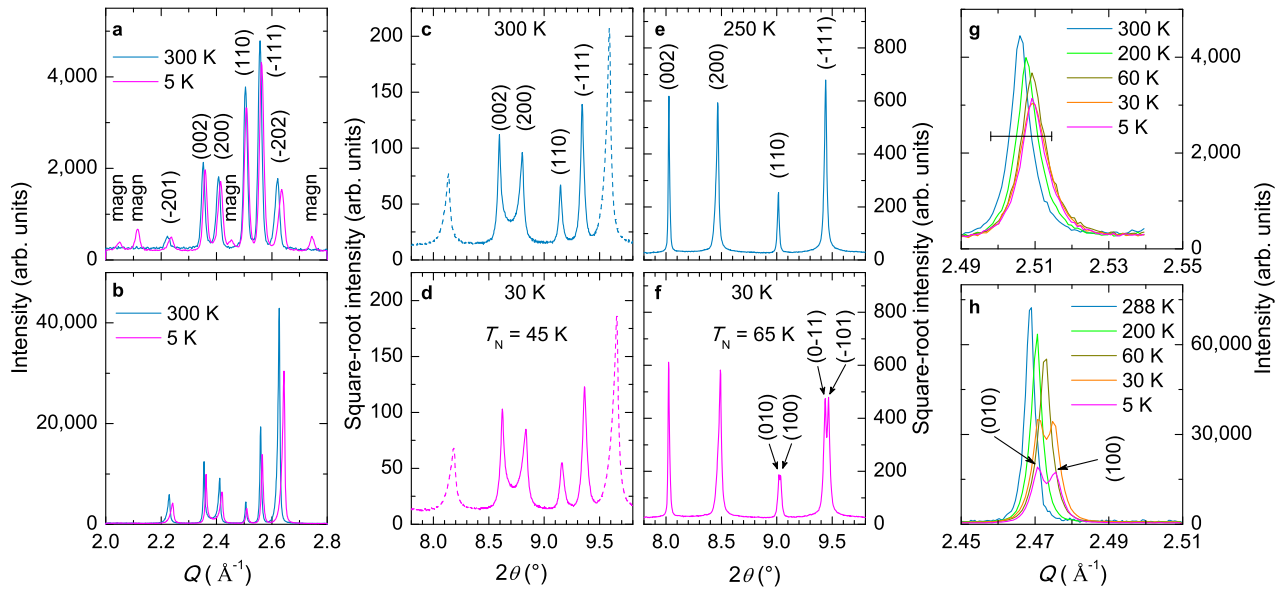}
\caption{{\bf Comparison of NPD and synchrotron XRD.} {\bf (a)} Neutron powder diffraction ($\lambda= 2.0787$~\AA) and {\bf (b)} synchrotron powder X-ray diffraction ($\lambda= 0.39986$~\AA) patterns of \alfa~at 5 and 300~K. Part of the synchrotron XRD diffraction patterns of {\bf (c,d)} \alfa~and {\bf (e,f)} the isostructural CuMnO$_2$, above and below the N\'eel temperature $T_{\rm N}$. In CuMnO$_2$ the characteristic peak splitting of certain families of Bragg reflections signifies a bulk phase transformation from the monoclinic to the triclinic structure. The temperature evolution of the (110) Bragg reflection from the synchrotron XRD pattern of {\bf (g)} \alfa~and {\bf (h)} CuMnO$_2$. The scale bar in {\bf (g)} indicates the full-width-at-half-maximum of the same Bragg reflection from NPD at 300~K. ($khl$) indices denote nuclear Bragg reflections and "magn" magnetic ones. arb. units, arbitrary units. The error bars in the diffraction patterns are defined as a square root of the counted neutrons/photons.}
\label{fig5}
\end{figure*}

The ID31 synchrotron experiment provides a significantly increased angular resolution (nominal instrumental contribution to the FWHM is $\sim0.003^\circ$ in $2\theta$), as compared to the previous NPD structural study \cite{Giot} (FWHM was $\sim0.3^\circ$ in $2\theta$).

\vspace{0.5 cm}
\noindent {\bf HAADF-STEM measurements.}
The structure of \alfa~at room temperature was examined by imaging a large number of nanocrystalline regions by HAADF-STEM. The images were obtained with the aberration-corrected Titan G3 electron microscope operated at 300 kV. The noise on the HAADF-STEM images was reduced by applying low-pass Fourier filtering.
 As \alfa~is air-sensitive, the TEM specimen was prepared in the Ar-filled glove box by further crushing the polycrystalline sample in a mortar in anhydrous ethanol or n-hexane and depositing drops of suspension onto holey carbon grids. The specimen was transported and inserted into the microscope under dry Ar, completely excluding contact with air.

\vspace{0.5 cm}
\noindent {\bf Modelling of the structure.}
Rietveld refinements of the crystal structure were performed with the FULLPROF suite \cite{J. Rodriguez-Carvajal}. The model included a pseudo-Voigt peak-shape function and the Stephens's formalism of anisotropic peak broadening \cite{Stephens}, in which the anisotropic strain broadening parameters $S_{hkl}$ measure the covariance $\langle \left(\alpha_i-\langle\alpha_i\rangle \right)\left(\alpha_j-\langle\alpha_j\rangle \right) \rangle$ of the distribution of metric parameters. The largest $S_{040}$ parameter thus measures the covariance $\langle \left(\alpha_2-\langle\alpha_2\rangle \right)^2 \rangle$ of the distribution of the metric parameter $\alpha_2=1/b^{*2}$.

The Rietveld refinement at 300~K shown in Fig.~\ref{fig2}a is in accord with the $C2/m$ monoclinic symmetry; $R$-factors not corrected for background are $R_{\rm p}=11.6\%$, $R_{\rm wp}=15.8\%$ and $R_{\rm exp}=3.80\%$, while $\chi^2=17.3$. The quality of fit with the same structural model is worse for $T<T_{\rm N}$ (at 30~K, $R_{\rm p}= 14.9\%$, $R_{\rm wp}= 18.9\%$, $R_{\rm exp}= 3.9\%$, $\chi^2= 23.1$) and the strain parameters are significantly increased. Surprisingly, the highly parametrised (as defined by the much larger number of generated reflections due to $\bar{1}$ Laue symmetry) strained $P\bar{1}$ monoclinic model that was suggested by the NPD study \cite{Giot} produces even poorer quality of fit ($R_{\rm p}= 17.2\%$, $R_{\rm wp}= 22.4\%$, $R_{\rm exp}= 3.9\%$, $\chi^2= 32.3$). Allowing the coexistence of strained monoclinic and unstrained triclinic phases, a significantly improved quality of fit ($R_{\rm p}= 8.7\%$, $R_{\rm wp}= 11.3\%,$ $R_{\rm exp}= 3.9\%$, $\chi^2= 8.2$), shown in Fig.~\ref{fig2}b, is attained, yielding a "parasitic" triclinic phase with a volume fraction of $\sim 35(1)\%$.

The influence of the planar defects detected by HAADF-STEM on the XRD pattern was simulated using a recursion algorithm realized in the DIFFaX program \cite{Treacy}. The model of the LiMnO$_2$ defect structure \cite{Croguennec} was initially used and adopted to mimic the coherent twin boundaries with a single \alfa~(-101) layer as the twin plane or stacking faults containing two layers. For the instrumental broadening characteristic for the ID31 beam-line already 2\% of faults provide noticeable anisotropic broadening, as shown in Fig.~\ref{fig2}f,g.

\vspace{0.5 cm}
\noindent {\bf Compatibility of the synchrotron XRD and NPD data.}
In the earlier NPD study \cite{Giot}, the conclusion that the nuclear triclinic cell with $P\bar{1}$ symmetry was required to describe the magnetic order, was reached. However, no direct evidence of the structural phase transformation from the monoclinic to the triclinic structure, i.e., peak splitting of particular families of nuclear Bragg reflections, was found. Complex structural modelling, such as that concluded in the present study, was impossible for the NPD data due to limited resolution. With inherent higher resolution of the synchrotron XRD data, on the other hand, symmetry-lowering peak splitting of nuclear reflections can be inspected and heavily parametrised two-phase structures can be witnessed.

The improved resolution capacity of the synchrotron XRD data is obvious from the comparison of representative diffraction patterns (in $Q$) from the NPD and synchrotron XRD studies at the same temperatures  (Fig.~\ref{fig5}a,b). In the NPD case we note the significantly broader nuclear Bragg reflections in the intermediate $Q$-range where instrumental resolution is optimal. Despite the increased angular resolution of the synchrotron XRD data the patterns of \alfa~below $T_{\rm N}$ do not show any resolvable peak splitting, in contrast to the isostructural CuMnO$_2$ (Fig.~\ref{fig5}c-f) \cite{Vecchini,Poienar}. In Fig.~\ref{fig5}g,h the temperature evolution of the representative (110) reflection from the high-resolution synchrotron XRD data of \alfa~is compared to that from CuMnO$_2$ that was obtained at very similar conditions. Even though the instrumental resolution is the same in both cases, the (110) reflection for the Na-based compound (Fig.~\ref{fig5}g) entails a larger FWHM at room temperature than that of the Cu-based analogue (Fig.~\ref{fig5}h). While it becomes progressively broader with decreasing temperature, it never shows a peak splitting below $T_{\rm N}$, in contrast to CuMnO$_2$.
\begin{figure}[t]
\includegraphics[trim = 1mm 5mm 4mm 20mm, clip, width=1\linewidth]{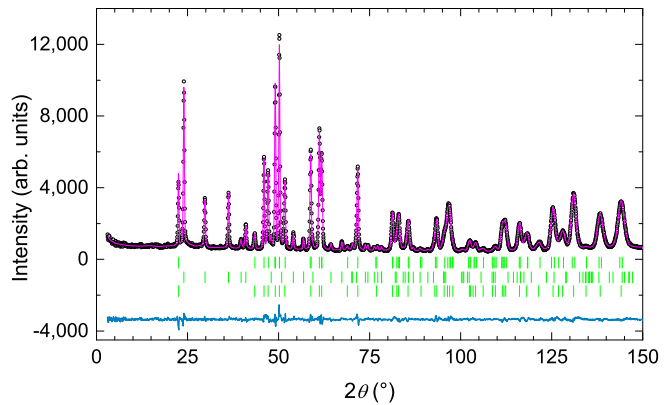}
\caption{{\bf Low-$T$ NPD profile of \alfa.} Observed (circles), calculated (thin lines) and difference (thick line) NPD profile at 5~K (BT1, NIST), after Rietveld refinement with the three-phase model number 3. Predicted reflection positions are marked by the vertical ticks for the triclinic nuclear [top set; $a= 3.1655(2)$~\AA, $b= 3.1575(2)$~\AA, $c= 5.7777(3)$~\AA, $\alpha= 110.395(4)^\circ$, $\beta= 110.292(5)^\circ$, $\gamma= 53.686(4)^\circ$], triclinic magnetic (intermediate set) and the monoclinic nuclear [bottom set; $a= 5.6495(2)$~\AA, $b= 2.8546(2)$~\AA, $c= 5.7835(2)$~\AA, $\beta= 113.036(5)^\circ$] structures. arb. units, arbitrary units. The error bars in the diffraction patterns are defined as a square root of the counted neutrons.}
\label{fig6}
\end{figure}

\begin{table*}[t]
\caption{{\bf Neutron powder diffraction modelling.} A Comparison of the quality of fit ($R$-factors) for various types of modelling by Rietveld analysis of the \alfa~NPD data at 5~K ($T< T_{\rm N})$. m, monoclinic and t, triclinic nuclear structures. magn, magnetic structure settings.}
\begin{center}
\begin{ruledtabular}
\begin{tabular}
{c c c c c c c} 
Number & Model
 & $\chi^2$ & $R_{\rm Bragg}$ (\%) & $R_{\rm magn}$ (\%) & $R_{\rm Bragg}$ (\%) & $R(F^2)$ (\%) \\
 & (Space Groups)
 &  & Monoclinic (m) & Magnetic &  Triclinic (t) & \\\hline 
 1 & $\rm m$+magn ($C2/m$, $C\bar{1}$) & 3.88 & 1.62 & 5.86 & N/A & 1.24 \\
  2 & $\rm t$+magn ($P\bar{1}$, $P\bar{1}$) & 2.60 & N/A & 3.96 & 2.10 & 1.34 \\  
  3 & $\rm t$+magn+$\rm m$ ($P\bar{1}$, $P\bar{1}$,$C2/m$) & 2.54 & 1.58 & 4.25 & 2.49 & 1.21
\end{tabular}
\end{ruledtabular}
\end{center}
\label{S1}
\end{table*}

In addition to the monoclinic (model number 1) and triclinic nuclear structures (model number 2; used in Ref.~\onlinecite{Giot}), the mixture of both structures (model number 3), utilized in the analysis of the synchrotron XRD data (Fig.~\ref{fig2}b), was tested against the NPD data where also the magnetic structure was taken into account. Fig.~\ref{fig6} shows the result of the 5-K NPD Rietveld analysis based on the model number 3. In a manner similar to the synchrotron XRD Rietveld analysis, the refinement claims the majority monoclinic phase ($\sim$85\%) that coexists with the minority phase entailing the nuclear and magnetic triclinic contributions. However, as summarized in Tab.~\ref{S1}, no significant overall improvement in the $R$-factors (quality of the fit) is achieved with the model number 3, suggesting that the NPD data are insensitive to the complexity of the phase-separation model. The NPD pattern is misleadingly overparametrised with the three-phase Rietveld fit, and as such, it does not carry the statistical accuracy to allow unambiguous refinement of the parameters from the model number 3.

The $R$-factors derived from the analysis of the NPD data are not directly comparable to those derived from the Rietveld analysis of the synchrotron XRD, because of quantitatively different nature structure factors. The paper on the NPD \cite{Giot} reports on the two-phase model entailing the nuclear ($R_{\rm Bragg}$) and magnetic structures ($R_{\rm magn}$), while in the case of the present synchrotron XRD study the $R$-factors were derived from the two-phase structural model (each phase has its own $R_{\rm Bragg}$). Furthermore, while the medium-resolution NPD data were significantly less sensitive to peak-shape parametrisation, in the high-resolution synchrotron XRD data the choice of the peak shape (pseudo-Voigt function), which is combined with the Stephens $hkl$-dependent anisotropic broadening may be a limiting factor that reflects in the magnitude of the $R$-factors.

\vspace{0.5 cm}
\noindent {\bf $^{23}$Na NMR measurement.}
Frequency-swept $^{23}$Na ($I=3/2$) NMR spectra were measured in the temperature range between 5 and 300~K on a home-built spectrometer equipped with a helium-flow cryostat in the fixed magnetic field of $B_0=8.9$~T. The spectra were obtained with a solid-echo pulse sequence. The $\pi/2$ pulse length and the interpulse delay were 3.4 and 30~\textmu s, respectively. Spin-lattice relaxation-rate ($1/T_1$) measurements utilized an inversion-recovery sequence at the peak of the central ($-\frac{1}{2}\leftrightarrow \frac{1}{2}$) nuclear transition. The NMR line shift was measured relative to the reference Larmor frequency $\nu_0=100.5255$~MHz, corresponding to the NMR signal of the 0.1~M NaCl solution.

The $^{23}$Na NMR spectra consist of a sharp central line ($-1/2\leftrightarrow 1/2$) and a background of satellite transitions ($\pm 1/2 \leftrightarrow  \pm 3/2$), broadened in first order by a quadrupolar coupling of the nuclear quadrupolar moment with the electric field gradient (EFG). The model Hamiltonian used for fitting the spectra included the Zeeman coupling to the applied magnetic field, the quadrupolar coupling and an uniaxial anisotropic transferred hyperfine coupling ${\underline A}_{jk}$ of the nuclear spin ${\bf I}_j$ with electronic spins ${\bf S}_k$ on nearby Mn$^{3+}$ sites, 
$\mathcal{H}^{\rm hf}_j= {\bf I}_j\cdot\sum_{k}{\underline{A}_{jk} \cdot {\bf S}_k}=\hbar \gamma {\bf I}_j\cdot\underline{\delta}\cdot {\bf B}_0$ ($\hbar $ is the reduced Planck constant and $\gamma=70.8$~MHz~T$^{-1}$ the $^{23}$Na gyromagnetic ratio). Isotropic powder averaging was performed. The obtained EFG asymmetry parameter is $\eta=0.54(3)$ and the quadrupolar frequency is $\nu_{\rm Q}=1.51(5)$~MHz, which is a rather typical value for the $^{23}$Na nuclei. The isotropic part of the hyperfine magnetic coupling tensor $\underline{\delta}_{\rm iso}$, which is responsible for the shift of the central line from the reference frequency $\nu_0$, is $\delta_{\rm iso}=707(10)$~ppm at 300~K. The anisotropic part, responsible for the line-shape of the central transition, has the principal eigenvalue $\delta_{\rm aniso}=950(10)$~ppm at 300~K. The quadrupolar interaction leads to the second-order shift of -20~ppm and the width of 100~ppm of the central line, which is small compared to the contribution of the hyperfine interaction. Also typical chemical-shifts of $^{23}$Na with the range -60 -- 10~ppm are negligible.

Magnetization-recovery curves in spin-lattice relaxation measurements were fitted to $M_{\rm z}(t)=a\left\lbrace1-\frac{1}{5}{\rm exp}\left[-(t/T_1)^\beta\right]-\frac{9}{5}{\rm exp}\left[-(6t/T_1)^\beta\right] \right\rbrace+b$, suitable for magnetic relaxation of the central line of a $I=3/2$ nuclei \cite{Suter}. Here, the parameter $\beta$ denotes a stretch exponent and $a,b$ are fit parameters.

\vspace{0.5 cm}
\noindent {\bf $\mu$SR measurements.}
$\mu$SR experiments were performed on the General Purpose Surface-Muon Instrument (GPS) at the Paul Scherrer Institute, Switzerland. Zero external magnetic field was applied and the temperature was varied between 2 and 300 K. Veto mode was utilized to minimize the background signal.

\vspace{0.5 cm}
\noindent {\bf Determination of the muon stopping sites.}
{\it Ab initio} calculations of the electrostatic potential in the monoclinic setting were carried out by applying the Quantum Espresso code \cite{Gianozzi}, and by using the generalized-gradient approximation (GGA) \cite{Perdew} for the exchange-correlation potential. The electron-ion interactions were described with the Vanderbilt ultrasoft potentials \cite{Vanderbilt}. The plane-wave cut-off parameters were set to 408 eV and 3264 eV, respectively, whereas a $4\times 4\times 4$ ($4\times 8\times 4$ for the non-spin-polarized case) mesh in the full Brillouin zone was used for the Methfessel-Paxton sampling \cite{Methfessel} integration. The self-consistency criterion was the total-energy difference of subsequent calculations not exceeding 10$^{-7}$ Ry.

It is supposed that $\mu^+$ would most likely stop at the global potential minimum of the crystal structure \cite{Yaouanc}, which has been verified before on several instances \cite{Luetkens, Maeter, Pregelj}. Both the spin-polarized and non-polarized calculations yield the same most-probable muon site $P_1=(0.18,0,0.47)$, which is an interstitial site with site symmetry $m$ at Wyckoff position 4$i$ in the monoclinic setting. Although this site is close to the Mn layer (see Fig.~\ref{fig4}b), its nearest-neighbour site is actually the oxygen site at the distance of 1.74~\AA. Calculations of the dipolar field at this site (\textit{vide infra}) yield the magnetic field of 0.24~T that nicely corroborates the experimental $B_{\mu 1}=0.22(2)$~T. However, our \textit{ab initio} calculation could not find any local minimum that would correspond to $B_{\mu 2}=0.69(2)$~T. Therefore, our assignment of the second site relies on a common tendency of the muon to "bind" to O$^{2-}$ at the distance of 1.0~\AA~\cite{Yaouanc}. We then set the second muon stopping site to $P_2=(0.45,0,0.75)$ (4$i$) on the 1.0~\AA-sphere around the oxygen site, where the dipolar field matches $B_{\mu 2}$ and the electrostatic potential of the unperturbed crystal structure is minimal. We stress that the sites $P_1$ and $P_2$ do not correspond to the two different structural phases of \alfa. Within these, the structural changes are minimal and could not result in the largely different internal fields.

The influence of a muon stopping at either $P_1$ or $P_2$ on all of the exchange constants and thus locally on the magnetic ground state is expected to be negligible. The site $P_1$ is far from any exchange path, while the site $P_2$ could possibly interfere only with the weak inter-layer path. Therefore, it is highly unlikely for the muon to cause any local inhomogeneities of the magnetic order.

\vspace{0.5 cm}
\noindent {\bf Modelling of the magnetic inhomogeneity.} 
Calculations of the dipolar magnetic field
${\bf B}_{\rm dd}({\bf r})=\frac{\mu_0}{4\pi}\sum_j \left\lbrace \frac{\bm{\mu}_j}{|{\bf r}-{\bf r}_j|^3}-\frac{3({\bf r}-{\bf r}_j)\left[\bm{\mu}_j\cdot({\bf r}-{\bf r}_j)\right]}{|{\bf r}-{\bf r}_j|^5} \right\rbrace$ 
at a given site ${\bf r}$ were performed by taking into account all Mn$^{3+}$ moments $\bm{\mathrm{\mu}}_j$ at sites ${\bf r}_j$, within a sphere centred at ${\bf r}$ with a radius large enough to ensure convergence. The moments were either taken polarized along the applied field in the paramagnetic state or ordered according to Ref.~\onlinecite{Giot} ($\mu=2.9~\mu_{\rm B}$ at 5~K). These calculations assumed a homogenous triclinic crystal structure, however, the results can be safely employed as the structural differences of the two phases are minimal and, therefore, do not noticeably affect the magnetic fields at the sites of interest. The dipolar field at the $^{23}$Na site at 300~K yields the anisotropic hyperfine coupling of $\delta_{\rm aniso}^{\rm dd}=\frac{A\chi}{N_{\rm A}}=1000$~ppm, which matches astonishingly well with the experimental $\delta_{\rm aniso}=950(10)$~ppm. Likewise, the local dipolar field at the site $P_1$ at 5~K, $B_{\rm P1}=0.24$~T, matches the experimentally observed field $B_{\mu 1}=0.22(2)$~T, which makes all other contributions to the local field negligible.

When modelling the normalized local magnetic-field distributions $D(B)=\frac{1}{N_0}\frac{{\rm d}N}{{\rm d}B}$ at a given site in the structurally inhomogenous phase below $T_{\rm N}$, we assumed that the magnetic order in the triclinic "parasitic" structural regions was inverted with respect to the monoclinic matrix 
as shown in Fig.~\ref{fig1}a, but all the moments were still pointing along the easy-axis direction set by a strong single-ion magnetic anisotropy \cite{Zorko}. 
The magnetic field at a chosen crystallographic site was calculated in $N_0=1024$ different crystallographic cells, for a given random distribution of defects, with their size characterized by the correlation length $\xi$ and their concentration yielding the fraction $p$ of all spin sites within the defect regions. Each set of $N_0$ points then yielded a particular field distribution $D(B)$. We considered one- (1D), two- (2D) and three-dimensional (3D) defects of size $\xi$. For the 1D defects this meant fractions of the $b$ chains of length $\xi$, for the 2D defects circles with diameter $\xi$ on the $ab$ planes, and for the 3D detects spheres with diameter $\xi$. Different concentrations $p$ and sizes $\xi$ of such defects were tested. The latter were either finite, being within the range of 5 to 70~\AA~for all three different dimensionalities, or infinite ($\xi=\infty$) for 1D and 2D collinear and planar defects, respectively.

For each model, the calculated local-field distributions $D(B_{\mu,j})$ at both muon stopping sites $j$ define the modelled muon relaxation function $P(t)=\frac{1}{3}+\frac{2}{3}\sum_{j=1}^2 f_j \int D(B_{\mu,j}){\rm cos} (\gamma_\mu B_{\mu,j} t) {\rm d}B_{\mu,j}$. 

\vspace{0.5 cm}
\noindent {\bf Calculation of the energies.}
The total energies were calculated {\it ab initio} by applying the Quantum Espresso code \cite{Gianozzi}. The generalized-gradient approximation (GGA) \cite{Perdew} was used for the exchange-correlation potential and the Vanderbilt ultrasoft potentials \cite{Vanderbilt} for describing the electron-ion interactions. The plane-wave cut-off parameters were set to 408 eV and 3264 eV, respectively, whereas a $8\times 16\times 8$ mesh in the full Brillouin zone was used for the Methfessel-Paxton sampling integration \cite{Methfessel}. The total-energy difference of subsequent calculations was limited below 10$^{-7}$ Ry for the self-consistency criterion.
In these calculations the position of individual nuclei was first relaxed by means of minimizing the total energy and the interatomic forces. 
Outside the calculation errors the monoclinic structure was found lower in energy than the triclinic structure for the nonmagnetic case, while the opposite 
relation was found for the antiferromagnetically ordered state.

The magnetoelastic contribution arises from the coupling between the 
strain-tensor components $\epsilon_{ij}$ ($i={\rm x,y,z}$) and the 
magnetization-direction vector ${\bf m}=(m_{\rm x},m_{\rm y},m_{\rm z})$. 
The respective energy-density contains the terms 
$b_{ij}\epsilon_{ij}m_i m_j$ \cite{duTremmolet}, 
where $b_{ij}$ denotes the 
magnetoelastic-coupling coefficients, which are a measure of the magnetoelastic contribution to
the total energy. The magnetization-direction 
dependence of the energy arises from the spin-orbit coupling that was taken 
into account for the Mn atoms. Individual coupling constants $b_{ij}$ can 
be extracted from linear fits to the calculated differences in the total energy
upon the magnetization-direction switching as a function of the appropriate
strain-tensor components (see, e.g., Ref.~\onlinecite{Komelj} for a detailed description of this procedure). The by far dominant 
magnetoelastic coupling constant $b_{\rm xy}=-4.1$~MJ/m$^3$ corresponds to the 
shear strain that is in fact responsible for the triclinic distortion. For the 
experimental shear strain $\epsilon_{\rm xy}=0.0023$, that results from the monoclinic-to-triclinic transformation
this gives the magnetoelastic energy of 2.5~$\mu$eV per unit cell.

\section*{Acknowledgements}
We are grateful to A.~Abakumov for the HAADF-STEM measurements and for useful discussions on the emerged physical crystallography. We thank A.~Amato for assistance with the $\mu$SR measurements and I.~Margiolaki for help with the ID31, ESRF experiments. Access to the synchrotron radiation facilities at ESRF, the NIST center for neutron research and to the facilities of the Integrated Infrastructure Initiative ESTEEM2 (EU 7th FP program with ref.~no.~312483) is gratefully acknowledged. We acknowledge the financial support of the Slovenian Research Agency (Project No.~J1-2118 and Program No.~P1-0125) and EU FP6 (Contract No.~011723-RICN).

\section*{Author contributions}
A.Z., D.A. and A.L. designed and supervised the project. The samples were synthesized by O.A. and A.L., who also performed the synchrotron X-ray diffraction and analysed the data. The NMR experiments were conducted and analysed by A.Z. and D.A. The $\mu$SR experiments were carried out by A.L. and analysed by A.Z. The {\it ab initio} calculations were performed by M.K. All authors contributed to the interpretation of the data and to the writing of the manuscript.

\section*{Additional information}
\noindent {\bf Competing financial interests.} 
The authors declare no competing financial interests.

\end{document}